\documentstyle[aaspp4,12pt]{article}
\begin{document}
\input psfig.sty

\title{Radial Color Gradient and Main Sequence Mass Segregation in M30
(NGC~7099)\footnote{Based on observations with the NASA/ESA {\it Hubble Space
Telescope\/}, obtained at the Space Telescope Science Institute, which is
operated by the Association of Universities for Research in Astronomy, Inc.,
under NASA contract NAS~5-26555.}}
\author{Justin H.\ Howell and Puragra Guhathakurta\footnote{Alfred P.\ Sloan
Research Fellow}}
\affil{UCO/Lick Observatory, Department of Astronomy \& Astrophysics,\\
University of California, Santa Cruz, California 95064, USA\\
Email: {\tt jhhowell@ucolick.org}, {\tt raja@ucolick.org}}
\and
\author{Amy Tan}
\affil{University of California, Davis, California 95616, USA\\
Email: {\tt aatan@ucdavis.edu}}

\begin{abstract}
It has long been known that the post--core-collapse globular cluster M30
(NGC~7099) has a bluer-inward color gradient, and recent work suggests that
the central deficiency of bright red giant stars does not fully account for
this gradient.  This study uses {\it Hubble Space Telescope\/} Wide Field
Planetary Camera~2 images in the F439W and F555W bands, along with
ground-based CCD images with a wider field of view for normalization of the
non-cluster background contribution, and finds $\Delta(B-V)\sim+0.3$~mag for
the overall cluster starlight over the range $2''$ to $\gtrsim1'$ in radius.
The slope of the color profile in this radial range is:
$\Delta(B-V)/\Delta\log(r)=0.20\pm0.07$~mag~dex$^{-1}$, where the quoted
uncertainty accounts for Poisson fluctuations in the small number of bright
evolved stars that dominate the cluster light.  We explore various algorithms
for artificially redistributing the light of bright red giants and horizontal
branch stars uniformly across the cluster.  The traditional method of
redistribution in proportion to the cluster brightness profile is shown to be
inaccurate.  There is no significant residual color gradient in M30 after
proper uniform redistribution of all bright evolved stars; thus the color
gradient in M30's central region appears to be due entirely to
post--main-sequence stars.  Two classes of plausible dynamical models,
Fokker-Planck and multimass King models, are combined with theoretical stellar
isochrones from Bergbusch \& VandenBerg (1992) and from D'Antona and
collaborators to quantify the effect of mass segregation of main sequence
stars.  In all cases, mass segregation of main sequence stars results in
$\Delta(B-V)\sim-0.06$ to $+0.02$ mag over the range $r=20''\>$--$\>80''$;
this is consistent with M30's residual color gradient within measurement
error.  The observed fraction of evolved star light in the $B$ and $V$ bands
agrees with the corresponding model predictions at small radii but drops
below it for $r\gtrsim20''$.
\end{abstract}


\section{Introduction}

Due to their high density, cores of globular clusters serve as excellent
laboratories for studying stellar interactions and the resulting changes in
stellar populations (\cite{hut92}; \cite{meyl97}).  In particular, a small
fraction of clusters known as post core collapse (PCC) clusters are
characterized by extremely high central densities and tend to have color
gradients in the sense of becoming bluer inward.  Though several authors have
shown that central depletion of bright red giant branch (RGB) stars is an
important contributor to the color gradient in these clusters
(cf.~\cite{piotto88}; \cite{burgbuat}; \cite{zoppr}, hereafter GWYSB), a
satisfactory explanation of the underlying physical cause of this RGB
depletion has proven elusive (\cite{djor91}).  Most globular clusters are
well fit by King models (KMs; \cite{king62}) and very few of these show any
such color gradient.  Even the KM clusters that are suspected to have a color
gradient (e.g., NGC~4147) are all quite centrally concentrated so that the
distinction between KM and PCC clusters is unclear in these cases
(\cite{djor92}, 1993).

M30 is a prototypical PCC cluster, with one of the best-studied color 
gradients of all globular clusters (\cite{willbahc}; \cite{chunfree};
\cite{cord}; \cite{peterson86}; Piotto et~al. 1988; \cite{burgbuat}; GWYSB).  
This gradient has traditionally been explained by a deficiency of red giants
and asymptotic giant branch stars near the cluster center, but Burgarella~\&
Buat (1996) found that this central deficiency does not account for the
observed color gradient, and GWYSB have independently suggested that the
central evolved star deficit only produces one third of the observed color
gradient in M30, and all evolved stellar populations put together produce
less than half of the observed gradient.  A possible explanation for the rest
of the color gradient is mass segregation of main sequence stars, since stars
near the main sequence turnoff, with higher mass and bluer color, are
expected to be more centrally concentrated than the fainter, redder, and less
massive stars.

This paper reexamines the radial color profile of M30's starlight using {\it
Hubble Space Telescope\/} ({\it HST\/}) Wide Field Planetary Camera~2 (WFPC2)
images and wider field ground-based images, tests methods for redistributing
the light of bright stars, and addresses the question of whether main
sequence mass segregation produces the necessary color gradient to explain
the observations.  Observations of M30's color gradient and methods for 
uniformly redistributing the light of bright evolved stars are described in
\S\,2.  Calculations of the effects of main sequence mass segregation are
presented in \S\,3, and \S\,4 contains a summary of the main points of the
paper.

\section{Observed Color Gradient}
\subsection{Non-Cluster ``Background'' Brightness}

In order to study the color of the cluster starlight, it is necessary to
determine and correct for the ``background'' level in the {\it HST}/WFPC2
images.  The term ``background'' includes all non-cluster contributions to
the total light, including foreground zodiacal light, extragalactic background
light, and even non-astrophysical artifacts such as low level residual cosmic
rays and hot pixels.  GWYSB estimated this background to be
0.094~ADU~pixel$^{-1}$ in F439W and 0.054~ADU~pixel$^{-1}$ in F555W based on
visually selected regions of the image more than $1\farcm5$ from the cluster
center and away from resolved stars.  This technique could be biased in
either direction: the background would be overestimated if unresolved cluster
starlight contributes significantly to the background over the entire image,
or underestimated if atypically dark regions were selected that
systematically avoid, for example, hot pixels, residual cosmic rays, and/or
background galaxies.

This paper uses a combination of {\it HST}/WFPC2 data and short ground-based
$B$- and $V$-band CCD exposures of M30 provided by Mike Bolte (see
\cite{bolte87}; \cite{Sandquist} for details) to determine the background
brightness in the WFPC2 images, in contrast to the method used by GWYSB.  The
ground-based images are used only for background estimation and not for
stellar photometry.  They cover a sufficiently large field of view that
unresolved cluster light is unlikely to affect measurements of the background
in the far corners of the CCD frames.  The background in these images
includes a dominant atmospheric airglow component in addition to the sources
listed above.  The mean $B$- and $V$-band background brightnesses are
measured in selected regions of these images located at projected distances
of $r\sim3'\>$--$\>4\farcm7$ and $r\sim8'\>$--$\>11'$, respectively, from the
cluster center; these background estimates are then subtracted from the
images.  Note, the ground-based $B$ data consist of two images, a core
pointing covering most of the WFPC2 field of view, and a southwest pointing
which overlaps the core pointing but not the WFPC2 field.  Since the core
image extends only to $r\sim3'$, the southwest image is used for background
estimation.  This background estimate is bootstrapped to the core image using
a difference image: core image minus registered, background-subtracted
southwest image.  The difference image is nearly free of star-subtraction
residuals for $r\sim90''\>$--$\>180''$; the mean value of regions selected
from this area, avoiding obvious residual artifacts, is used as the
background value for the core image.  The $V$-band background measurement is
more straightforward and is based on a single CCD image.  The regions used to
measure the mean non-cluster background flux in the southwest $B$ and $V$
images are selected to be away from all resolved stars ($B<16$, $V<19$) as
these are likely to be cluster members according to the \cite{ratbah}
Galactic star count model (see discussion at the end of this section).  The
$B$ and $V$ background estimates show no trend with angular separation from
the nearest bright star or from the cluster center, indicating that the
measurements are unaffected by scattered light from bright stars or
unresolved faint stars in M30.

The {\it HST}/WFPC2 PC1 and WF2--WF4 images are combined into a mosaic image
in each of the F439W and F555W bands (GWYSB).  These mosaic images are
rotated, rebinned, and gaussian smoothed to match the orientation and
resolution of the corresponding ground-based images ($B$:
$0\farcs6$~pixel$^{-1}$, $\rm FWHM\sim1\farcs5$; $V$:
$0\farcs44$~pixel$^{-1}$, $\rm FWHM\sim1\farcs7$).  The background-subtracted
ground-based images are then masked to preserve only the region of overlap
with the WFPC2 mosaic, taking care to mask out the edges of both images where
the smoothing of the latter is imperfect.  The ground-based $V$ image covers
all of the WFPC2 F555W image, while the $B$ image covers about 93\% of the
WFPC2 F439W image, excluding only a small corner section of WF2 at
$r\gtrsim1'$ from the cluster center.  The resulting {\it HST\/} images
differ from the corresponding ground-based images only in terms of a
photometric scale factor and the background flux in the WFPC2 images.

The WFPC2 background flux in each band is obtained in two different ways.
The first method involves a linear least squares fit to solve simultaneously
for the photometric scale factor and WFPC2 background flux level by doing a
pixel-to-pixel comparison of the $10^4\>$--$\>10^5$~pixels in each pair of
matched WFPC2 and ground-based images.  Because of its high stellar surface
density, coupled with inaccuracies in the PSF match, the PC1 CCD is not used
in this comparison.  The WFPC2 background flux is estimated to be 0.0083~ADU
in F439W and 0.092~ADU in F555W per $0\farcs0996\times0\farcs0996$ mosaic
image pixel.  Unless otherwise mentioned, these least-squares-fit--based
WFPC2 background values are adopted for the rest of the paper.  In the
second, more direct, method of background determination, the photometric
scale factor is estimated using aperture photometry on a selection of bright
stars which are known to be isolated on the original, high-resolution {\it
HST\/} images.  After applying this multiplicative photometric correction to
the ground-based image, a difference image (WFPC2 minus ground-based) is
constructed whose mean value should be equal to the WFPC2 background flux.
In practice, point spread function (PSF) matching and star subtraction are
not perfect, so it is necessary to estimate the mean background level in
regions away from bright star residuals.  A comparison of the two background
estimates and errors in the background flux determination are discussed in
\S\,2.2 and \S\,2.3.

The adopted WFPC2 background levels correspond to sky brightnesses of
$\mu_{\rm sky}(B)=24.84$~mag~arcsec$^{-2}$ and $\mu_{\rm
sky}(V)=21.45$~mag~arcsec$^{-2}$, but it should be noted that these
background brightness levels may be partly instrumental in origin---the Space
Telescope Science Institute pipeline bias subtraction may have been
inaccurate seeing as the data were obtained only a few months after the
installation of the WFPC2 instrument.  Spatial variations in sky brightness
due to individual Galactic field stars are expected to be unimportant.  For
example, the \cite{ratbah} Galactic star count model prediction in this part
of the sky is less than 1~star~arcmin$^{-2}$ with $V<20$ at the bright end of
the stellar distribution, which corresponds to a brightness level of
$\mu(V)\sim29$~mag~arcsec$^{-2}$.  Since Galactic stars tend to outnumber
distant field galaxies for $V\lesssim21$ (cf.~\cite{bgs}), the effect of
stochastic variations in the bright end of the field galaxy population is
likely to be even smaller.

\subsection{Color Profile of the Cluster Starlight}

This study adopts the eight~radial bins defined in GWYSB for the purpose of
studying M30's $B-V$ color profile.  The radial bins are chosen such that
each contains approximately the same number of evolved stars with
$V\lesssim19$, a sample dominated in number by faint RGB stars.  These bins
are:
(1)~$r<5\farcs00$,
(2)~$5\farcs00\leq{r}<9\farcs80$,
(3)~$9\farcs80\leq{r}<15\farcs41$,
(4)~$15\farcs41\leq{r}<23\farcs2$,
(5)~$23\farcs2\leq{r}<35\farcs8$,
(6)~$35\farcs8\leq{r}<51\arcsec$,
(7)~$51\arcsec\leq{r}<71\arcsec$, and
(8)~$71\arcsec\leq{r}<130\arcsec$.
The characteristic radius adopted for each of these bins, for the purpose of
comparison with model predictions, is the median radial distance of the stars
in that bin (Table~\ref{table2}).  As described in GWYSB, the total $B$ and
$V$ flux within each radial bin is derived from direct aperture photometry on
the background-subtracted WFPC2 F439W and F555W mosaic images, respectively,
by differencing successive concentric circular apertures.  The contribution
of specific stellar populations, on the other hand, is determined by summing
over the list of detected stars for which photometry has been carried out
using standard techniques (PSF fitting, aperture photometry).  Only bright
RGB and horizontal branch (HB) stars are analyzed in this manner for the
purpose of uniform redistribution (\S\,2.3).  Detailed image simulations
indicate that the detection rate for these bright stars is practically 100\%
(\cite{yanny}; \cite{gysb96}) so that it is not necessary to apply any
radially-dependent incompleteness correction.

The mean $B-V$ color of M30 within each radial bin (annulus) is plotted in
the top panel of Figure~\ref{cgrad} (long dashed line with bold triangles).
Monte Carlo realizations, based on a synthetic $V$-band luminosity function
and $B-V$ color distribution to mimic the properties of M30's stellar
distribution, show that Poisson fluctuations in the bright RGB and HB
population result in a random error of $\delta(B-V)=0.084$~mag for each
radial bin.  There is a significant radial gradient of about $+0.3$~mag from
the cluster center out to $r\sim1'$, followed by an abrupt change of about
$-0.2$~mag from $r=1'$ to $r=1\farcm5$.  Specifically, a least squares fit of
a straight line to the $B-V$ vs log($r$) profile results in a best-fit slope
of $0.13\pm0.06$ using all eight radial bins, and $0.20\pm0.07$ using only
radial bins~1--7 ($r\lesssim1'$).  These measurements are in keeping with
earlier observations of M30's central blueing trend (cf.~\cite{cord};
\cite{peterson86}; Piotto et~al.\ 1988; \cite{burgbuat}).  Note, the revised
WFPC2 background flux estimates (\S\,2.1) result in a color profile that is
somewhat different from that published by GWYSB using the same data set,
especially for $r\gtrsim40''$.  The open circles and horizontal bars show the
ground-based color measurements presented by Peterson (1986).  Peterson's
measurements in concentric circular apertures are converted to colors in
annuli by differencing successive apertures; note, these measurements are
drawn from complete annuli, while the WFPC2 image geometry forces radial bins
to have incomplete azimuthal coverage beyond $r\gtrsim15''$.  Nevertheless,
our {\it HST}-based and Peterson's ground-based measurements of the color
profile are in good agreement.  The slight overall difference in $B-V$ color
($\approx0.05$~mag in the mean) between the two data sets could be the result
of systematic differences in photometric calibration.

\begin{figure}
\plotone{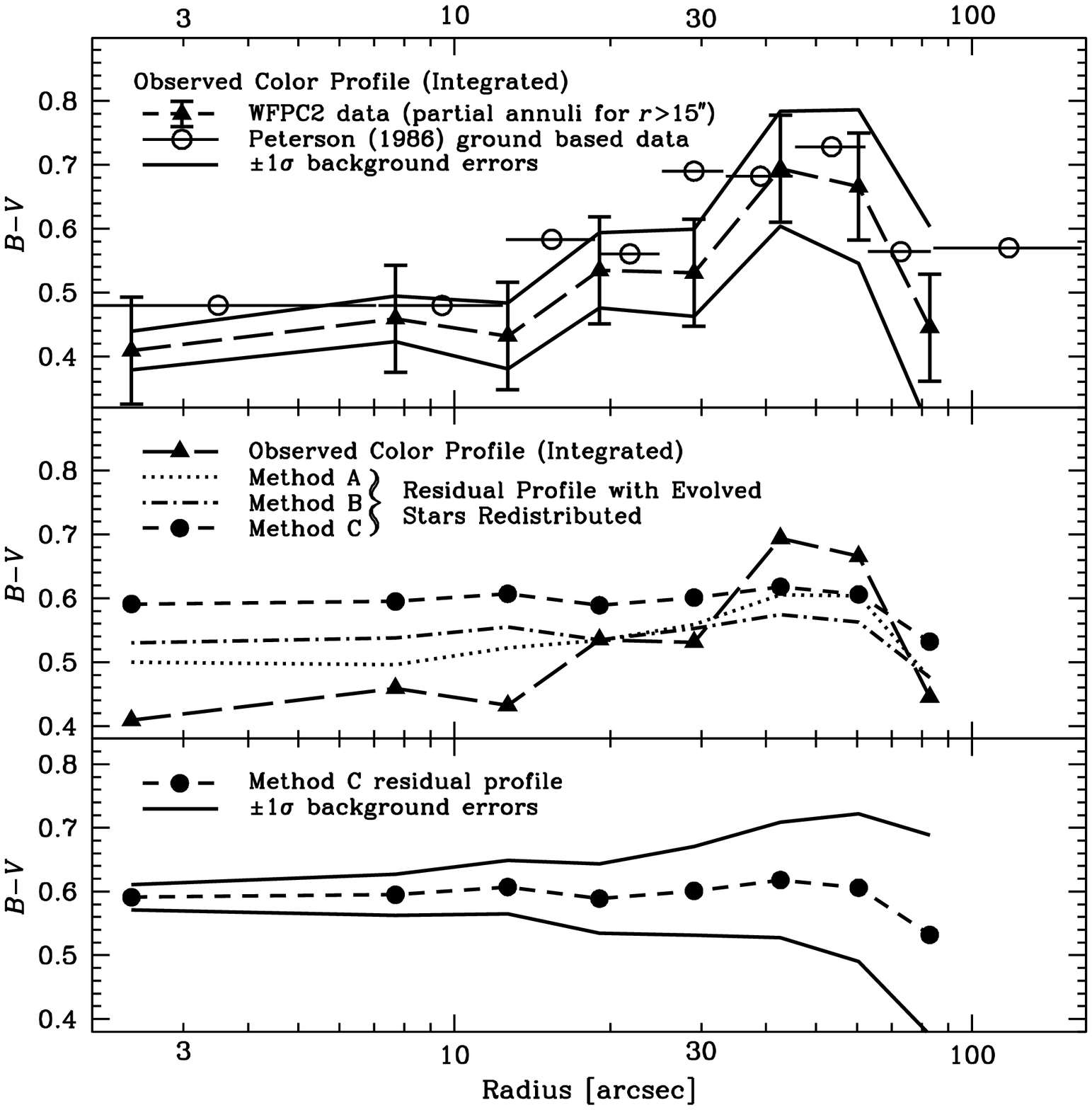}
\end{figure}
\begin{figure}
\caption{{\it Top panel}:~The average $B-V$ color of
M30's starlight, integrated over radial bins (annuli), as a function of radius
using a combination of {\it HST}/WFPC2 and ground-based images (long dashed
line with filled triangles).  The error bars account for Poisson variations
in the number of bright stars in each radial bin.  The solid lines illustrate
the effect of $\pm1\sigma$ conservative error estimates in the background
flux level as described in \S\,2.2.  Open circles with horizontal bars show
the ground-based color measurements made by Peterson (1986).~~~~
{\it Middle panel}:~Residual $B-V$ color profiles after uniform redistribution
of the light of bright red giant branch and horizontal branch stars using
three different algorithms---~~~
Method~A:~proportional to overall cluster $B+V$ light as in GWYSB (dotted
line);
Method~B:~proportional to faint RGB stars using overall bright to faint RGB
ratio (dot-dashed line);
and Method~C:~proportional to faint RGB stars using bright to faint RGB ratio
in radial bins~5--8 (short dashed line with filled circles).
Method~C is the most accurate redistribution algorithm; see \S\,2.3 for a
complete discussion of the various redistribution schemes.  The long dashed
line and filled triangles represent the color profile of the cluster prior to
redistribution (same as upper panel).  The original color profile is jagged,
becoming redder by about $+0.3$~mag from the center out to $r\gtrsim1'$,
while redistribution of HB and bright RGB flux results in a smooth profile
that is consistent with no residual color gradient.~~~~
{\it Bottom panel}:~The Method~C redistributed color profile (short dashed
line with bold circles, as in the middle panel), with the effect of
$\pm1\sigma$ conservative background errors on the redistributed color
profile illustrated by the solid lines.
\label{cgrad}}
\end{figure}

The effect of the errors in background flux measurements on the final color
profile is calculated in two ways:

\begin{itemize}

\item[$\bullet$]{The formal error estimate is derived from the least
squares fit to the matched WFPC2 and ground-based pixels in the WF CCDs only:
the error in the mean background flux over this area is transformed to the
error per WFPC2 mosaic pixel by multiplying by $\sqrt{N_{\rm WFPC2}(\rm
3WF)}$, where $N_{\rm WFPC2}(\rm 3WF)\approx3\times800\times800$ is the
number of mosaic pixels in the area of the fit (3WF).  In the photometric
scale of the WFPC2 data set, the formal errors are 3.1~ADU in F439W and
2.0~ADU in F555W per mosaic pixel.  The effect of WFPC2 background error on
each radial bin is then computed by scaling the pixel-to-pixel error by the
square root of the number of mosaic pixels in that radial bin.}

\item[$\bullet$]{The conservative error estimate is obtained by
performing an independent least squares analysis on each of the three WF
CCDs.  The variation in the mean background flux amongst the three WF CCDs is
substantially larger than the formal error in the mean given by the fitting
routine.  This variation is likely a result of systematic errors arising from
PSF mismatch and/or bias subtraction differences from CCD to CCD.  Radial
bin~7 has an area comparable to that of a single WF CCD, so a conservative
error for bin~7 is derived from the spread of mean background values amongst
the three WF CCDs: the spread in the mean is scaled by the number of pixels
in each fit (the number of matched pixels in each WF CCD) to convert the
error in the mean to an error in the sum, and multiplied by the square root
of the ratio of the area of bin~7 to that of a WF CCD,
$\sqrt{A({\rm radial~bin~7})/A({\rm WF})}$.  The background errors for the
other radial bins ($i=1,8$) are estimated in similar fashion by multiplying
by $\sqrt{A({\rm radial~bin}~i)/A({\rm WF})}$.  The conservative error
estimate for radial bin~7 is (practically) free of the assumption that the
error in the background flux scales as the square root of the area since
$\sqrt{A({\rm radial~bin~7})/A({\rm WF})}=0.93$.  This area ratio factor
departs most strongly from unity for the innermost bins, but then the
background flux (and the error in the background) is a negligible fraction of
the total flux in these bins.}

\end{itemize}

The error in the measurement of the background level in the ground-based
images is determined from the variance amongst mean values in different blank
regions.  The variance is scaled to the area of each radial bin and is added
in quadrature to the WFPC2 background error (formal and conservative) for
that bin.  Note, the uncertainty in the background level of the ground-based
images is generally unimportant in relation to the uncertainty in the WFPC2
background; the ground-based background error is only 20\% of the overall
error in the case of the formal $B$ band error and even less in the other
cases.  The net $B$- and $V$- band background errors are combined in order to
assess the effect on the color profile of M30's starlight; the $\pm1\sigma$
conservative error in the background is shown by the solid lines in the top 
panel of Figure~\ref{cgrad}.

The formal error is about an order of magnitude smaller than the conservative
error; these estimates probably represent lower and upper bounds,
respectively, on the true error.  In assuming $\sqrt{A}$ scaling of the
summed background flux, the formal error estimate ignores systematic errors
such as PSF mismatch and possible CCD-to-CCD differences in residual bias
level.  The conservative error estimate, on the other hand, is derived from a
comparison of the three~WF CCDs which are known {\it a priori\/} to have
different stellar densities and hence different degrees of systematic error
caused by PSF mismatch of bright stars.  The direct measurement of the
background in bright-star--free regions agrees with the background value
derived from the least squares fit to the full area of the three WF CCDs to
well within the conservative error (\S\,2.1).

\subsection{Uniform Redistribution of the Light of Evolved Stars}

M30's nucleus has been shown to be deficient in the most luminous RGB stars.
The ratio of the surface density of bright RGB stars to the cluster surface
brightness (Piotto et~al. 1988) or to the surface density of faint
RGB/subgiant stars (GWYSB) is significantly lower in the central $r<30''$
than further out in the cluster.  The central four radial bins ($r<23''$)
contain only 23~bright RGB stars, less than 40\% of the expected number (60).
The expected bright RGB number is derived from the observed number in radial
bins~5--8 ($23''<r<130''$) which are unaffected by the central deficiency of
these stars, normalizing to the faint RGB population; a comparable bright RGB
fraction is estimated from Peterson's (1986) ground-based data in the
$r=1'$--$3'$ region of the cluster.  Similarly, the net bright RGB flux in
the inner four radial bins is 40.9\% of the flux in radial bins~5--8.  Is
the central deficiency of bright RGB stars entirely responsible for M30's
bluer-inward color gradient?  The light from bright RGB stars must be
redistributed in some {\it uniform\/} way in order to determine how much of
the observed color gradient is due to this central bright RGB depletion.

In GWYSB, bright RGB flux was redistributed following the radial dependence
of the total $B+V$ flux, with the relative normalization between the
two~bands chosen such that faint RGB stars (with the same color as the
average cluster color) contribute equally in $B$ and $V$.  The result is the
dotted curve in the middle panel of Figure~\ref{cgrad}; we refer to this as 
Method~A.  While this traditional method of bright RGB redistribution
algorithm is convenient to implement, it is obviously inaccurate.  In the
limiting case where the population being redistributed has a negligible
contribution to the overall cluster light, Method~A produces uniform
redistribution.  In general, however, the stellar population of interest
accounts for a finite fraction of the total light, so that the redistributed
color profile contains a ``ghost'' of the original color profile.  This is
due to the fact that the normalizing function for redistribution---the
integrated cluster light profile---is itself influenced to some degree by the
original radial distribution of the stars being redistributed.  Specifically,
redistributing the bright RGB stars in M30 by Method~A results in a larger
residual bluer-inward color gradient than the more accurate methods defined
below (Fig.~\ref{cgrad}), because bright RGB stars contribute 30\%--50\% of
the total cluster light.

A more reasonable scheme might be to redistribute the bright RGB light in
proportion to the faint RGB stars.  For example, this may be achieved by
assigning one eighth of the total bright RGB flux to each radial bin
(dot-dashed curve in middle panel of Fig.~\ref{cgrad}; Method~B), since the
radial bins were defined to contain approximately equal numbers of faint RGB
stars.  Note, however, that the total bright RGB flux in the WFPC2 image of
M30 is lower than it would have been in the absence of the central bright RGB
depletion.  Making the assumption that the relative abundance of the bright
RGB population is `normal' beyond $r\sim25''$ in M30, the preferred method of
redistribution is to assign one fourth of the bright RGB flux summed over
radial bins~5--8 to each of the 8~radial bins, again roughly in constant
proportion to faint RGB stars (Method~C; short dashed line and solid circles
in middle and bottom panels of Fig.~\ref{cgrad}).  Unlike the other
redistribution methods, Method~C actually adds in extra bright RGB flux to
compensate for the central bright RGB deficiency in the cluster.  If there is
no residual color gradient, this simply has the the effect of producing a
redward color offset with respect to the Method~B profile.  However, if a
residual color gradient is present, Method~C produces both a shallower color
gradient and redder overall colors than Method~B because of dilution by the
extra bright RGB light.  The two new bright RGB redistribution methods used
in this paper, Methods~B and C, are coupled with redistribution of the flux
of HB stars: each radial bin is assigned one eighth of the total HB flux,
thereby correcting for Poisson noise in the distribution of HB stars and
forcing the HB to faint RGB flux ratio to be constant.  Table~\ref{table1}
summarizes the redistribution methods described above and lists both the
formal and conservative error estimates.  The solid lines in the bottom panel
of Figure~\ref{cgrad} show the effect of $\pm1\sigma$ conservative background
errors on the Method~C residual profile.

\begin{figure}
\vskip -2.0truein
\centerline{\psfig{figure=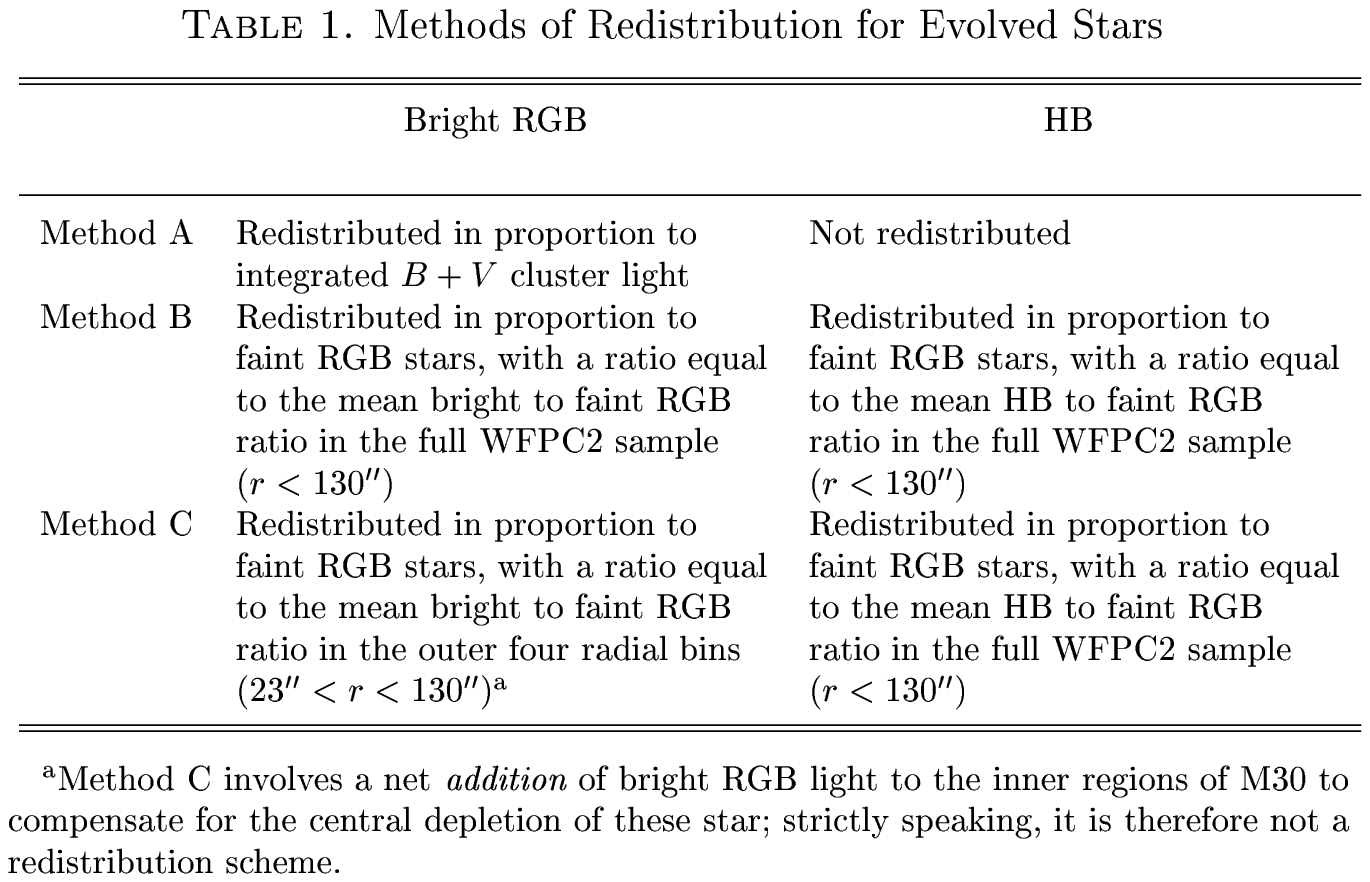}}
\end{figure}
\begin{table}\dummytable\label{table1}\end{table}

The motivation for these redistribution schemes is that mass segregation is
{\it not\/} expected to produce radial gradients in either the bright to
faint RGB ratio or the HB to faint RGB ratio, as the characteristic masses of
faint and bright RGB stars differ by only $\Delta{M}\lesssim0.03\,M_\odot$
(\cite{bergbvan}), and the typical masses of HB stars are thought to be
within $0.1\,M_\odot$ that of faint RGB stars (\cite{ldz90}).  Moreover, the
dynamical timescale for mass segregation is about a factor of~4 longer than
the lifetime in the HB evolutionary phase (\cite{djor93a}; \cite{lee90}).

Redistribution of both bright RGB and HB flux results in a residual color
profile that is consistent with no residual color gradient, and quite smooth
compared to the observed color profile (Fig.~\ref{cgrad}).  The jagged nature
of the original color gradient results from relatively small numbers of
bright RGB and HB stars dominating the light at any given radius.  The
smoothness of the residual, redistributed color profile indicates that the
photometry and subtraction of these bright evolved stars must be accurate.
The slight kink in radial bin~3 may be due to oversubtraction at the level of
1--2\%.  A $\chi^2$ test shows that a constant color (no gradient) is an
adequate fit to the Method~C residual color profile, and no significant slope
is found.

\begin{figure}
\plotfiddle{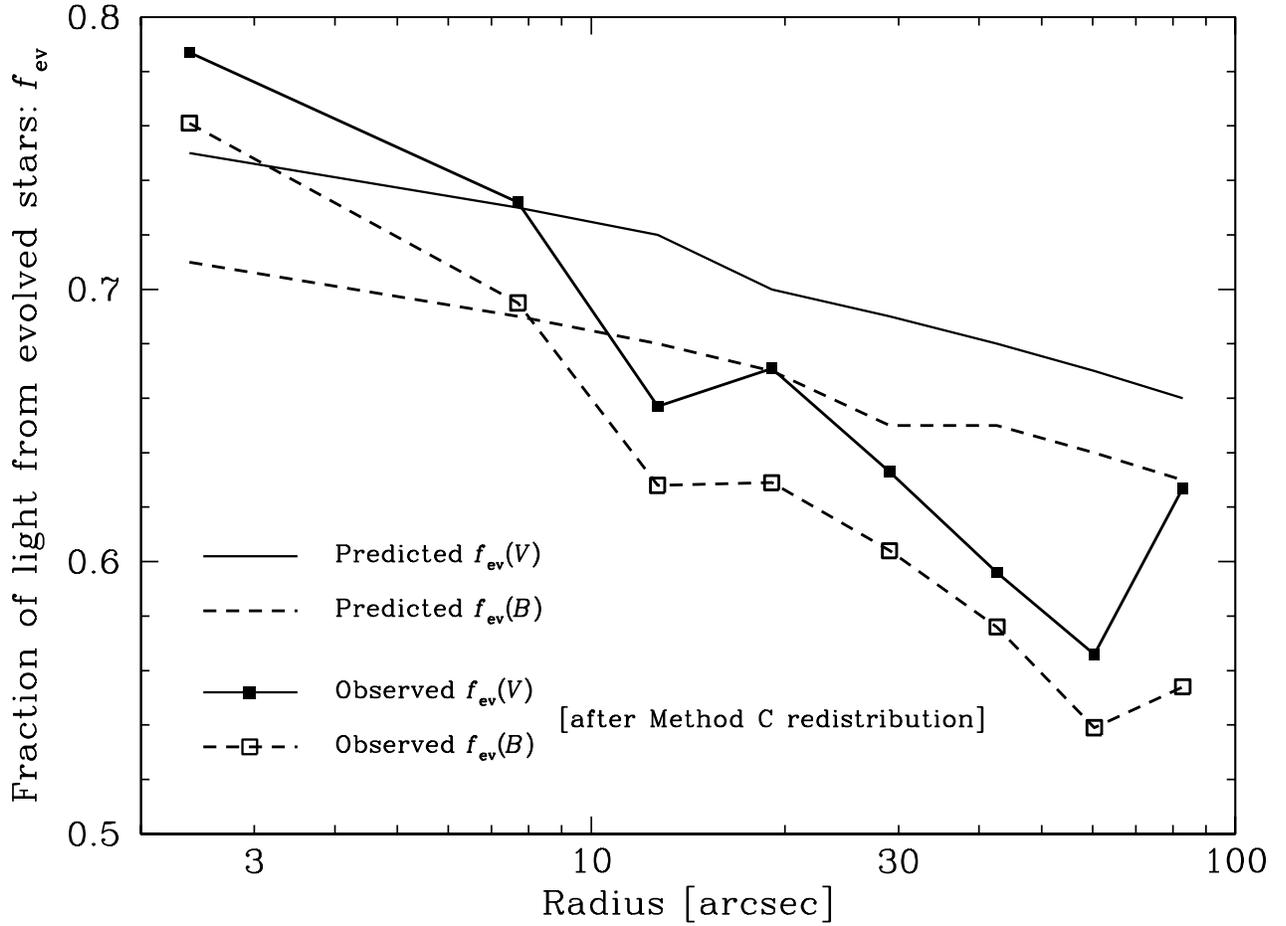}{4.0truein}{-90}{64}{64}{-243}{402}
\caption{The ratio of the light of evolved stars
($V<18.6$) to the light of all stars, $f_{\rm ev}$, as a function of
radius in M30.  The `observed' $f_{\rm ev}$ values in the $B$ and $V$ bands
represent measurements made after the bright RGB and HB stars have been
redistributed via Method~C (dashed line with open squares and solid line with
filled squares, respectively).  Predictions based on the
Bergbusch~\& VandenBerg (1992) 10~Gyr theoretical isochrones are shown as
dashed and solid lines (without symbols) for the $B$ and $V$ bands,
respectively.  The redistributed evolved star fraction in both bands
falls progressively below the theoretical prediction with increasing radius
beyond $r\sim10''$.
\label{fev}}
\end{figure}

The fraction of $B$- and $V$-band light from evolved stars, defined to be
those with $V<18.6$ (bright/faint RGB, HB, subgiants, blue stragglers), is
measured in each radial bin after uniform redistribution of bright RGB and HB
stars using Method~C.  These fractions, $f_{\rm ev}(V)$ and $f_{\rm ev}(B)$,
are shown in Figure~\ref{fev} (bold and open squares) and listed in
Table~\ref{table2}.  Evolved stars contribute about three quarters of the
total flux at small radii in both bands and their flux fraction falls off
with increasing radius to about~0.6 at $r\gtrsim1'$.  The $f_{\rm ev}(V)$
fractions are needed for appropriate normalization of the models in \S\,3.3;
the observed $f_{\rm ev}(V)$ and $f_{\rm ev}(B)$ values are also compared to
fractions derived from the Bergbusch~\& VandenBerg (1992) theoretical
luminosity function (\S\,3.4).  Although $f_{\rm ev}$ is calculated after
evolved star redistribution, these values will be referred to as ``observed''
$f_{\rm ev}$ values for the rest of this paper.

\section{Effect of Main Sequence Mass Segregation}
\subsection{Dynamical Models}

A Fokker-Planck (FP) dynamical model (\cite{dull}) designed specifically for
the cluster M15, constrained by its measured velocity dispersion, surface
brightness profile, and millisecond pulsar acceleration, is used in this
paper.  No FP model has been designed for M30, so the M15 model is adapted to
M30 by applying a small adjustment to its radial scale (B.\ W.\ Murphy 1998,
private communication).  Additional procedures used to tailor the model to
M30 are described in \S\,3.3.  There are important differences between the
physical parameters of M15 and M30 some of which may even be relevant to the
central depletion of bright red giants---e.g.,~M15's central velocity
dispersion and central density are about twice the corresponding M30 values
(\cite{dubath97}; \cite{prymey93}).  However, the FP model is merely used to
characterize mass segregation in M30, recognizing that the model is not
expected to apply to M30 in detail.  The effect of varying the nature and
degree of mass segregation is explored in \S\S\,3.3--3.4.

The FP model specifies the number of stars in each of twenty stellar mass
bins as a function of radius.  The first five mass bins represent nonluminous
stellar remnants (neutron stars and white dwarfs of various masses) which are
irrelevant for the color gradient computation.  The next two mass bins
correspond to HB and RGB stars, respectively, and the last thirteen mass bins
cover main sequence stars of successively lower mass in the range
0.74$\>$--$\>0.11\,M_\odot$.  The model mass functions are not strictly
monotonic with stellar mass at all radii, though they display a general
increase in number of stars towards smaller stellar mass, more so at large
radii than at small radii.

As an alternative to the FP dynamical model, a multimass KM is also
considered.  Although PCC clusters are thought to be in a nonequilibrium
state, multimass KMs provide an adequate description of the observed
variation in mass function slope as a function of radius
(cf.~\cite{sosinking}).  \cite{bolte89} derived the mass function slope at
$r\approx100''\>$--$\>400''$ in M30 from studies of the luminosity function
of faint main sequence stars, and fit the data with a multimass KM from
\cite{pryor86} using an assumed core radius of $10''$ (see Fig.~3 of Bolte
1989).  In the present study, this KM is used to predict the mass function
slope $x$ as a function of radius in the inner part of M30: $x=-2.75$ in
radial bin~1 ($r\sim2''$) and $x=+0.25$ in radial bin~8: ($r\sim80''$).  Note
$x$ is defined in the usual way: $dN(M)\propto{M}^{-(1+x)}dM$.

\subsection{Stellar Evolution Models}

Theretical stellar isochrones provide the luminosity and color of stars as a
function of their mass.  We use a Bergbusch~\& VandenBerg (1992) isochrone
with $\rm [Fe/H]=-2.03$, $\rm [O/Fe]=+0.70$, and $\rm Y=0.235$, consistent
with the estimated abundance of M30 (cf.~\cite{zinnwest}; \cite{carretta};
\cite{Sandquist}).  The calculations presented in this paper are based on a
10~Gyr isochrone; the main result, however, is insensitive to choice of
isochrone age.  Bergbusch~\& VandenBerg compute evolutionary tracks in
$(M_V,~B-V)$ space as well as stellar luminosity functions for a variety of
power law initial mass function slopes.

Recently, D'Antona and collaborators have calculated stellar evolutionary
tracks using updated input physics (P.~Ventura 1998, private communication).
The models are based on the assumption of gray atmospheres for stellar masses
$M>0.6\,M_\odot$ (\cite{dantona}) and use model atmospheres from \cite{hab99}
and \cite{al00} for $M<0.6\,M_\odot$ (\cite{montalban}).  The \cite{castelli}
conversion from effective temperature to $B-V$ color is used in both mass
ranges.  Our study uses the 10~Gyr, $\rm [Fe/H]=-2$ isochrone from D'Antona's
group; this is hereafter referred to as the ``D'Antona'' theoretical
isochrone.  The lower main sequence portion of the D'Antona isochrone
($M_V>9$) is bluer by up to $\Delta(B-V)=-0.2$~mag than the Bergbusch~\&
VandenBerg (1992) isochrone.  Moreover, stars of a given mass are brighter by
as much as $-0.5$~mag in $M_V$ in the former isochrone.  These differences
have a negligible impact on the predicted overall $B-V$ color profile
(\S\,3.3; Table~\ref{table2}).

\subsection{Predicted Color Gradient}

The dynamical models and stellar isochrones described above are used to
compute the $B-V$ color of the cluster light as a function of radius.  For
computational convenience and by analogy with the M30 data set, the dynamical
models are normalized to a fixed number of red giants in each of the 8~radial
bins.  The radii in the dynamical models are compared directly to the
observed (projected) radii in M30; projection effects are ignored since M30's
radial brightness profile is relatively steep.  If anything, projection tends
dilute the model color gradient.  Whereas, as we show below, the color
gradient predicted by mass segregation models is already very small.

In the case of the FP dynamical model, mass bin~\#7 represents red giants and
contains stars with mass $M\sim0.8\,M_\odot$ covering a wide range of
absolute magnitudes: $M_V\approx-2.7$ (tip of RGB) to $M_V\approx+3.6$ (main
sequence turnoff).  What is the appropriate stellar luminosity to attach to
this mass bin?  A direct flux-weighted integration of M30's observed evolved
star luminosity function in the $V$ and $B$ bands over the range $M_V<+3.6$
yields a characteristic absolute magnitude of $(M_V)_{\rm RGB}=+1.2$ and a
color of $(B-V)_{\rm RGB}=+0.80$ for the set of red giants, subgiants and
turnoff stars represented by mass bin~\#7.  The HB stars in the FP dynamical
model (mass bin~\#6) are tailored specifically to M15, and M30 has a
significantly different HB morphology and HB to RGB ratio.  Each HB star is
assigned an absolute magnitude of $(M_V)_{\rm HB}=+0.45$ and a color of
$(B-V)_{\rm HB}=-0.05$, as measured for M30's short blue HB.  The total
number of HB stars predicted by the FP model across all 8~radial bins is
scaled so that the overall HB/RGB $V$-band flux ratio matches that observed
in M30 (after correcting for central bright RGB depletion), while preserving
the FP model's run of HB/RGB ratio with radius (monotonic increase outward).
Main sequence stars in mass bins~\#8--\#20 are assigned $B$ and $V$
luminosities based on the Bergbusch~\& VandenBerg (1992) evolutionary tracks.

As a further adaptation of the FP model to M30, the number of evolved stars
(RGB+HB, mass bins~\#6 and \#7) is adjusted with respect to the number of
main sequence stars (sum of mass bins~\#8--\#20).  The adjustment is carried
out independently in each radial bin to ensure that the fractional evolved
star $V$-band flux, $f_{\rm ev}(V)$, in the model agrees with the `observed'
value in that radial bin of M30, after uniform redistribution of RGB and HB
stars (\S\,2.3; Table~\ref{table2}).  This method of normalizing to the
observed $f_{\rm ev}(V)$ values is preferred over a direct integration of the
model because M30 is observed to have a 30\% higher RGB-to-turnoff ratio than
predicted by models (\cite{Sandquist}; GWYSB).  At each radius, the overall
luminosities in the $B$ and $V$ bands are obtained by integrating over stars
in all the model mass bins.  This yields the predicted $B-V$ color of M30 as
a function of radius (``FP'' entry in Table~\ref{table2}).

\begin{figure}
\vskip -2.0truein
\centerline{\psfig{figure=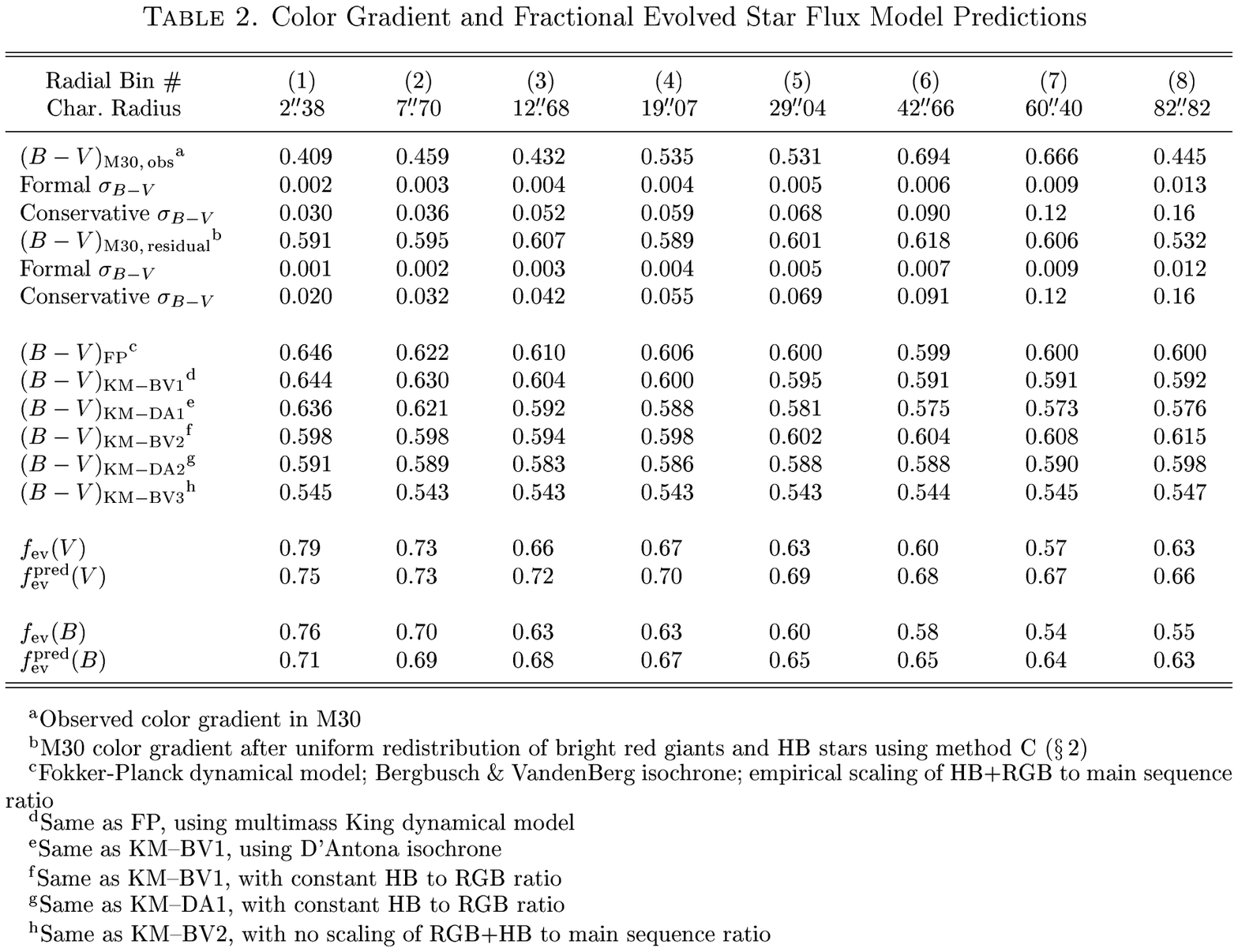}}
\end{figure}
\begin{table}\dummytable\label{table2}\end{table}

The above technique is repeated for the multimass KM, using power law stellar
mass functions whose slope $x$ increases with radius (\S\,3.1).  The
calculation is performed using both sets of evolutionary tracks to attach a
$V$-band luminosity and $B-V$ color to stars of a given mass and the results
are listed in Table~\ref{table2}: ``KM--BV1'' (Bergbusch~\& VandenBerg 1992)
and ``KM--DA1'' (D'Antona).  As described above, the $f_{\rm ev}(V)$
normalization constraint (ratio of evolved stars to main sequence stars) is 
applied at each radius; HB stars are normalized to the RGB population
overall, and the radial dependence of the HB to RGB ratio is adopted from the
FP model.

To be consistent with the redistribution of HB stars in M30 and following the
reasoning given in \S\,2.3, calculations are also carried out in which the
ratio of model HB to RGB flux is forced to be constant with radius and equal
to the ratio measured in M30 (after bright RGB and HB redistribution).  The
calculations are otherwise identical to the ``KM--BV1'' and ``KM--DA1''
calculations described above.  The constant HB-to-RGB results are labeled
``KM--BV2'' and ``KM--DA2'' in Table~\ref{table2}.

The effect of relaxing the $f_{\rm ev}(V)$ normalization constraint is also
explored.  This is done via direct integration of the luminosity functions
tabulated by Bergbusch~\& VandenBerg (1992) for mass function slopes in the
range $x=0$ to +2.5.  The mass functions must be extrapolated to obtain the
negative slopes, $x\gtrsim-3$, appropriate for M30's central regions
(\S\,3.1).  The Bergbusch~\& VandenBerg model isochrones do not include the
HB phase of stellar evolution.  As in the ``KM--BV2'' calculation, HB flux is
added in constant proportion to RGB stars at all radii.  The resulting $B-V$
color gradient is labeled ``KM--BV3'' in Table~\ref{table2}.  Note, the
$f_{\rm ev}^{\rm pred}(V)$ values for the Bergbusch~\& VandenBerg isochrone,
with HB stars added in, tend to be higher than the observed values in M30 for
$r\gtrsim10''$ (with bright RGB and HB stars redistributed): the discrepancy
is about 10\% at $r\sim30''$ (Table~\ref{table2}; Fig.~\ref{fev}).

Comparing the results of the six calculations described above
(Table~\ref{table2}), it is clear that choice of dynamical model has little
effect on the color profile:  ``FP'' and ``KM--BV1'' model colors differ by
less than $\pm0.01$~mag at all radii.  The bluer lower main sequence in the
D'Antona isochrone relative to the Bergbusch~\& VandenBerg (1992) isochrone
(\S\,3.2) results in slightly bluer overall colors, though the color gradient
is nearly the same (``KM--BV1'' vs ``KM--DA1'' and ``KM--BV2'' vs
``KM--DA2'').  Lower main sequence stars produce a larger fraction of the
total light at large radii than at small radii, and this results in a
slightly greater difference in color between D'Antona and Bergbusch~\&
VandenBerg calculations at large radii.  Comparing the ``KM--BV1'' results to
``KM--BV2'' and ``KM--DA1'' to ``KM--DA2'' shows that the increase in HB/RGB
ratio with increasing radius in the former set of calculations is responsible
for the overall bluer colors at large radii; the constant HB/RGB ratio in the
latter set of calculations yields a marginally redder-outward gradient.
Unlike the other calculations, the ``KM--BV3'' case avoids empirical
normalization of the evolved-to-main sequence flux ratio.  Thus, ``KM--BV3''
has a higher evolved star fraction than the other cases since the
Bergbusch~\& VandenBerg $f_{\rm ev}^{\rm pred}(V)$ values are generally
higher than the observed $f_{\rm ev}(V)$ values; this results in slightly
bluer overall colors due to the increased prominence of evolved stars, most
notably bright blue HB stars.

A back-of-the-envelope calculation based on the ``KM--BV3'' model illustrates
why main sequence mass segregation can make no appreciable contribution to
M30's overall color gradient.  The mean cluster color of about
$\langle{B-V}\rangle_{\rm M30}=+0.7$, which also happens to be the mean color
of faint RGB stars, is a reasonable value at which to separate the main
sequence into an upper and lower main sequence (UMS and LMS, respectively).
Integrating the LMS, UMS, and evolved star portions of the isochrone shows
that the relative numbers of stars are: $N_{\rm LMS}\approx{N}_{\rm
UMS}\approx8\,N_{\rm ev}$ in radial bin~2; and $N_{\rm LMS}\approx\,4{N}_{\rm
UMS}\approx50\,N_{\rm ev}$ in radial bin~7.  Evolved stars contribute about
70\% of the total light in radial bin~2, and 55\% in radial bin~7.  The
typical evolved star (RGB/HB) is $\approx100\times$ more luminous in $V$ and
$\approx0.3$~mag redder in $B-V$ than typical UMS stars, and
$\approx1000\times$ more luminous and $\approx0.3$~mag bluer than LMS stars.
Thus, the color shift relative to the faint RGB due to UMS stars is the
product of their fractional flux contribution ($0.7\times0.01\times8$) and
the color difference ($-0.3$~mag): $\Delta(B-V)=-0.017$~mag in radial bin~2,
and $\Delta(B-V)=0.55\times0.01\times(50/4)\times(-0.3)=-0.021$~mag in radial
bin~7.  Similar calculations for the LMS yield shifts in $\Delta(B-V)$ of
$+0.002$~mag and $+0.008$~mag, respectively.  The estimated net color
gradient due to main sequence mass segregation from this admittedly
oversimplified calculation is
$\Delta(B-V)=-0.021+0.008-(-0.017+0.002)=+0.002$~mag redder outward, well
within the range of $+0.01$~mag to $-0.05$~mag yielded by more precise
calculations, and equal to the value of $+0.002$~mag from the most closely
related such calculation, ``KM--BV3'' (Table~\ref{table2}).

\subsection{Fraction of Evolved Star Light}

The calculation involving direct integration of the Bergbusch~\& VandenBerg
(1992) stellar mass and luminosity functions, ``KM--BV3'', predicts a larger
fraction of evolved star light for $r\gtrsim10''$ than the $f_{\rm ev}$
values observed in M30 (\S\,2.3).  Put another way, a higher fraction of main
sequence light is observed than predicted, especially as one moves away from
the cluster center.  This discrepancy is similar in $B$ and $V$.  To resolve
the discrepancy in the context of power law mass functions (as the multimass
KMs are parameterized above), the exponent must vary with radius from $x=-5$
in the innermost radial bin ($r\sim2''$) to $x=0$ around $r\sim20''$ to
$x=+1.5$ in the outermost radial bin ($r\gtrsim1'$).  However, the stellar
mass function is not necessarily a power law and there may be alternate ways
of resolving the $f_{\rm ev}$ discrepancy.  Also, the true discrepancy is
nearly 30\% greater than Figure~\ref{fev} indicates: the ``KM--BV3''
calculation does not take into account M30's 30\% higher ratio of RGB to
turnoff stars relative to model isochrones (\cite{Sandquist}; GWYSB).

As described in \S\,3.1, the mass function slopes used for the KM
calculations in this paper are derived from mass function measurements at
$r\gtrsim2'$ in M30 (Bolte 1989).  This model-based inward extrapolation is
sensitive to the choice of cluster core radius.  Recent high resolution
studies of the central region have shown that the effective cluster core
radius is significantly smaller than Bolte's assumed value of $r_{\rm
core}=10''$ (\cite{yanny}).  This would imply a somewhat stronger radial
dependence of the mass function slope $x$ over the $r\lesssim2'$ region than
we have adopted.  The KM--BV3 calculations are repeated using values of
$x=-5$ in the first radial bin to $x=0$ in the outermost radial bin, where
this range of $x$ values is obtained by a simple linear extrapolation of
Bolte's data points.  As the models of Pryor et~al.\ (1986) flatten out at
$x\gtrsim-3$ at small radii, this linear extrapolation is an unrealistically
extreme case.  The resulting color gradient is $\Delta(B-V)=-0.007$ mag bluer
outward, and, like the other calculations, is consistent with the M30's
residual color profile.  The more extreme mass segregation invoked to explain
the discrepancy between predicted and observed $f_{\rm ev}$ values ($x=-5$ to
+1.5; see previous paragraph) likewise has little net effect on the color
gradient.

The fractional degree of contamination by faint red foreground field stars
is expected to increase radially outward---e.g.,~the area of radial bin~1 is
79~arcsec$^2$ while that of radial bin~8 is 8440~arcsec$^2$, even though both
bins contain roughly the same number of cluster giants.  However, the density
of M30 stars is so high in the central region of the cluster covered by the
WFPC2 image that field star contamination should be negligible even in radial
bin~8.  The number of field stars predicted by the Galactic star count model
of \cite{ratbah} is too low by several orders of magnitude to have a
significant effect on M30's color profile.

\section{Conclusions}

The radial $B-V$ color profile of the post core collapse cluster M30 is
measured using {\it Hubble Space Telescope\/} Wide Field Planetary Camera~2
images, along with ground-based images whose wider field of view allows for a
reliable determination of the non-cluster background in the WFPC2 image.  M30
displays a significant radial color gradient of $\Delta(B-V)\sim+0.3$~mag,
corresponding to a slope of
$\Delta(B-V)/\Delta\log(r)=0.20\pm0.07$~mag~dex$^{-1}$ from
$r=2''\>$--$\>1'$.  An accurate new technique is developed for uniform
redistribution of the light of the brightest cluster stars, which compensates
for stochasticity in their spatial distribution and for the central depletion
of bright red giants.  There is no significant residual color gradient after
bright star redistribution, implying that post--main-sequence stars are
entirely responsible for the central color gradient in M30.  This is contrary
to the recent results of GWYSB and Burgarella \& Buat (1996), but confirms
the earlier finding of Piotto et~al.\ (1988).

The physical mechanism responsible for the central depletion of bright red
giants (and hence the color gradient) in M30 and in other post--core-collapse
clusters remains a mystery.  Direct stellar collisions are too infrequent to
destroy an appreciable fraction of the giants within their short lifetimes.
The lack of a comparable central depletion amongst horizontal branch stars,
the downstream evolutionary products of bright red giants, suggests a `short
circuiting' of the bright red giant phase rather than complete destruction of
such stars; this may bear some relation to the evolution of giants in binary
systems.  The reader is referered to \cite{djor91} and GWYSB and to
references therein for a detailed discussion of these issues.

This study also investigates the effect on the color profile of mass
segregation of main sequence stars in the context of cluster dynamical models
and theoretical stellar isochrones.  The model calculations show a slight
bluer-outward color gradient when the HB varies as predicted by the
Fokker-Planck dynamical model [$\Delta(B-V)\sim-0.06$ from
$r=20''\>$--$\>80''$], and an even smaller redder-outward gradient if the HB
is held constant with respect to the RGB.  In all cases, the color gradient
predicted by mass segregation models is consistent with the data to within
the measurement uncertainties.  The predicted fraction of light from evolved
stars using the theoretical mass and luminosity functions of Bergbusch~\&
VandenBerg (1992) suggests that there is a 10\%--30\% achromatic excess of
faint star light at large radii in M30 relative to conventional mass
segregation models.

\bigskip
\bigskip
\acknowledgments

We would like to thank Brian Murphy for providing an updated electronic
version of the Fokker-Planck models described in Dull et~al. (1997), Paolo
Ventura for providing the isochrones computed by D'Antona's group, and Mike
Bolte for providing ground-based images.  We are grateful to Sandy Faber and
Mike Bolte for a critical reading of the manuscript, and to the referee,
George Djorgovski, for several insightful comments.  PG would like to thank
his collaborators on the M30 WFPC2 study, Zo Webster, Brian Yanny, Don
Schneider, and John Bahcall, for useful discussions about M30's color
gradient in the context of an earlier paper that served as the motivation for
this work.  This project was supported in part by an undergraduate McNair
Scholarship from the University of California at Davis (AT).

\clearpage
\newpage

\end{document}